\documentclass{aa}
\usepackage{graphicx}
\usepackage{times}
\usepackage{astron}

\begin{document}

\thesaurus{3         
              (09.01.2; 
	       11.09.1 NGC 1569; 
               11.09.2; 
               11.09.5; 
	       11.11.1; 
               11.19.3)} 

\title{A low-mass HI companion of NGC1569?}

\author{J.M. Stil\inst{1}
\and F.P. Israel\inst{1}}
\institute{Leiden Observatory, PO Box 9513, NL 2300 RA Leiden, The Netherlands}

\offprints{J.M. Stil}
\mail{stil@strw.leidenuniv.nl}

\date{Received 4 March 1998; accepted 3 June 1998}
\maketitle

\begin{abstract}
High-sensitivity maps of the large-scale structure of atomic hydrogen in 
the starburst dwarf galaxy NGC 1569 show evidence for an HI cloud 
with a mass of $\rm 7 \cdot 10^6 M_\odot$, at a projected distance of 5 kpc 
from the parent galaxy. This cloud may be a condensation in a low-column-density 
HI halo or a companion galaxy/HI-cloud. NGC 1569 and its companion are connected 
by a low surface brightness HI bridge. At the edge of NGC1569, the HI bridge coincides 
with H$\alpha$ arcs, also detected in soft X-rays.

\end{abstract}

\section{Introduction}

NGC 1569 is a small Im type galaxy at a distance of $\rm 2.2 \pm 0.6\ Mpc$ 
(Israel 1988), and a probable member of the IC 342 group. Due to its low
galactic latitude, the members of this group are heavily obscured. The number
of known members is growing rapidly with the discovery of Dwingeloo 1
(Kraan-Korteweg et al. 1994), Dwingeloo 2 (Burton et al. 1996), MB 1 and 2
(McCall \& Buta 1995) and MB 3 (McCall \& Buta 1997).  

With a diameter of 1.85 kpc, NGC 1569 is dominated by the aftermath of a burst of 
star formation (e.g. Israel \& de Bruyn 1988, Vallenari \& Bomans 1996). It
suffers an outflow of gas (Waller 1991, Heckman et al. 1995 
and references therein) and contains three very luminous, compact star
clusters and several bright HII regions (Waller 1991). 

The nearest neighbour to NGC~1569 is the dwarf irregular galaxy UGCA~92 at a 
projected distance of 60 kpc 
(Karachentsev \& Tikhonov 1993), 
at approximately the same radial velocity ($v_{hel}=-99\ \rm km s^{-1}$), and with 
an HI mass of $\rm 7 \cdot 10^7 M_\odot$ (de Vaucouleurs et al. 1991). 
Emissionline-dominated dwarf galaxies such as NGC 1569 
often, if not always, have nearby companions (Taylor et al. 1995, but see 
Taylor et al. 1996). The pair NGC 1569/UGCA 92 fits the selection criteria 
and overall properties given by Taylor et al. (1995), but the orbital timescale of 
this pair seems far too long to explain the recent starburst in NGC 1569
in terms of interaction with UGCA 92.

The overall HI distribution consists of a high-column-den\-sity ridge with 
three peaks and an `arm' extending to the west
(Reakes~1980, Israel \& Van Driel 1990). 
Channel maps presented by Reakes (1980) also show weak emission south-east 
of the bright HI ridge, likewise visible in the low resolution map of 
Israel \& Van Driel (1990). 
In this Letter, we concentrate on the nature of this emission and compare it
with the extended HI haloes observed in the dwarf galaxies IC 10 (Shostak \& Skillman 1989)
and NGC4449 (Bajaja et al. 1994). 
A more complete analysis of the HI emission of NGC1569 is deferred
to a forthcoming paper.
    
\section{Observations and data reduction.}

HI in NGC 1569 was observed with the WSRT during four 12-hour runs
between November 1989 and January 1990. A total of 160 baselines ranging 
from 36 m to 2772 m were sampled at increments of 18 m. Because of missing
short interferometer spacings, structures larger than a few arcminutes 
become increasingly diluted. However, the HI emission in individual 
velocity channels generally does not extend that far, so that the 
importance of this effect is minor.

The 2.5 MHz band was sampled by 128 independent frequency channels of 4.12 km 
s$\rm ^{-1}$ width. The Hanning-smoothed velocity 
resolution is 8.2 km s$\rm ^{-1}$.
Maps at $\rm 27''$ and $\rm 60''$ resolution were 
produced by applying Gaussian weight functions of 843 m and 390 m width in 
the UV plane. 
The continuum was subtracted in each pixel 
by fitting a first-order polynomial to 48 line-free channels at both edges
of the frequency band. In each channel map,
the area containing line emission was identified manually and
CLEANed down to half the r.m.s. noise. 
The rms noise in each $\rm 27''$ channel map is 1.1 mJy per beam. The channel maps 
have not been corrected for primary-beam attenuation.

\section{Results and analysis.}

\begin{figure*}
\resizebox{\textwidth}{!}{\includegraphics[angle=-90]{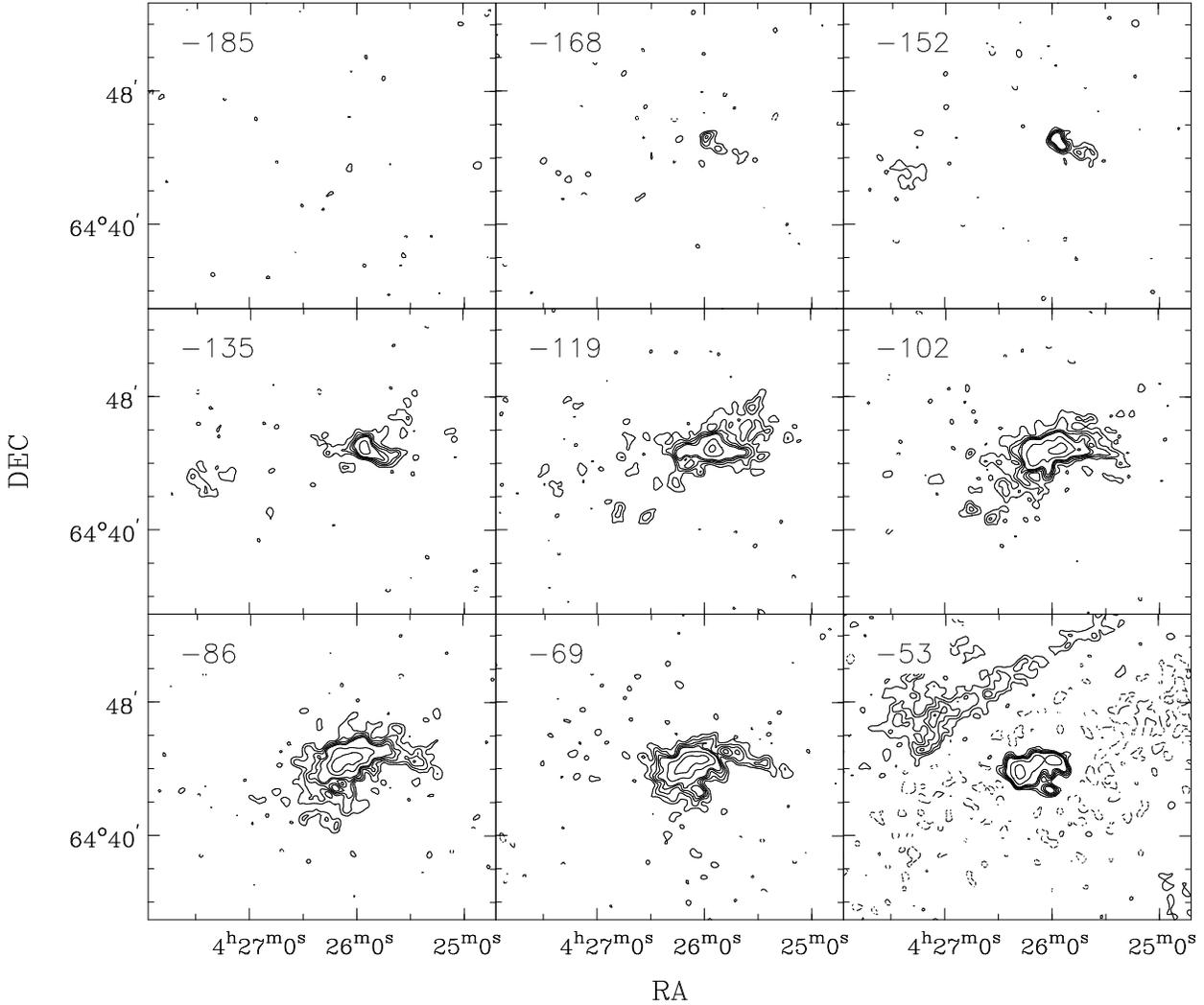}}
\hfill
\parbox[b]{\textwidth}
{
\caption{Velocity maps at $\rm 27''$ resolution, summed over four channels
each. Heliocentric velocities marked are those of the third channel in each 
set. Contours are at -10.5, 10.5 (3$\sigma$), 17.5, 24.5, 31.5, 38.5, 77.0 
and 154.0 mJy per beam. The last panel also contains Galactic foreground
emission present in a velocity range of about 10 $\rm km s^{-1}$.
}
\label{bigchanmaps}
}  
\end{figure*}

\begin{figure*}
\resizebox{\hsize}{!}{\includegraphics[angle=0]{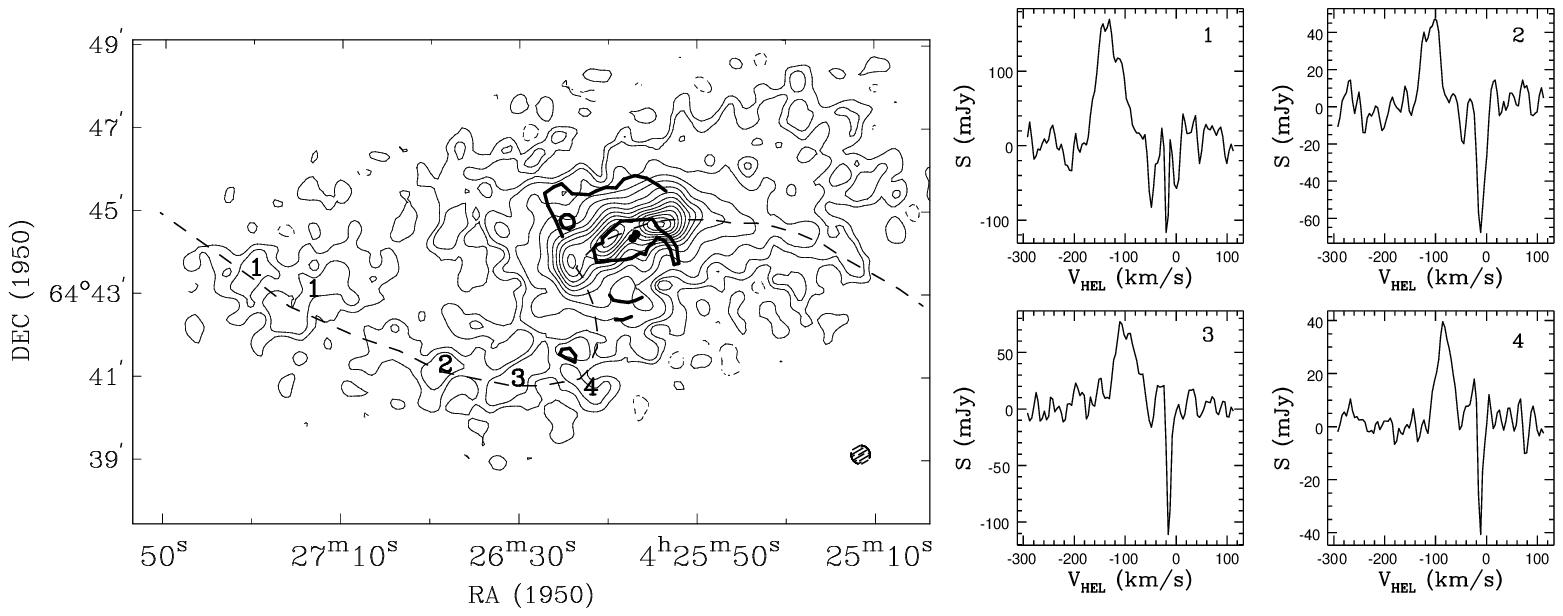}}
\caption{ HI column density map of NGC1569. Contours are at -1 
(dashed), 1, 2, 4, 6, 11, 16, 21, ...  $\cdot$10$^{20}$ HI $\rm cm^{-2}$.
The luminous star clusters A and B in the centre of NGC1569 are marked by dots. 
Thick lines mark the brightest $\rm H_\alpha$ emission, $\rm H_\alpha$ arcs and the
northern limit of filaments in a deep image by Hunter et al. (1993). The dashed curve
shows the path of the position-velocity diagram of Fig.~\ref{bridgeXV}. 
The synthesized beam is shown as a circle in the lower right corner.
{\it b).} Line profiles of the areas labeled 1 through 4 in {\it a}, defined 
by the $\rm 2 \cdot 10^{20}$ $\rm cm^{-2}$ contour. The line profile centroid 
shifts from -130 $\rm km\ s^{-1}$ to -80 $\rm km\ s^{-1}$ along the
HI bridge. Much narrower negative features are
sidelobes of Milky Way foreground emission (see Fig.~\ref{bigchanmaps} and 
Fig.~\ref{bridgeXV}).
} 
\label{profig}
\end{figure*}

In the velocity range $\rm -160\ km\ s^{-1}$ to $\rm -120\ km\ s^{-1}$, the
channel maps show emission from NGC 1569 itself, and also from
a second feature at $\rm \alpha=4^h 27^m 20^s$, $\rm \delta=64^o 43'.0$, best
identified in the $v_{hel}=-152$ and $v_{hel}=-135 \rm km\ s^{-1}$ maps in 
Fig.~\ref{bigchanmaps}. Although the emission is weak (4 $\sigma$), 
we consider it real. First, its location coincides with
emission in Reakes's independent maps. Second,
we eliminated the possibility of solar interference by excluding all 
hour angles with the sun above the horizon. We also extracted the line 
profile in the direction of the feature from the raw data, i.e. before
continuum subtraction and cleaning. In both cases, the emission persisted.
We will refer to this feature as NGC~1569-HI. There is further weak 
emission between NGC 1569 and NGC~1569-HI in the velocity range 
$\rm -130\ km\ s^{-1}$ to $\rm -90\ km\ s^{-1}$ (also visible in 
Reakes's $\rm -100\ km\ s^{-1}$ channel map. 

A second filamentary structure is visible in the channel maps with
velocities -86 and -69 
$\rm km\ s^{-1}$ near $\alpha=\rm 4^h 25^m 30^s$, $\delta=\rm 64^o 44'$.
It appears as a westward extension of the brightest emission. It is considerably 
closer to NGC~1569 than NGC~1569-HI and always connected to the brightest 
emission. For this reason, we treat it as a part of NGC~1569 and  
refer to it as the `western arm'. This feature was first noticed by Reakes (1980). 

A map of the total HI column density is shown in Fig.~\ref{profig}, 
together with line profiles summed over selected regions. The emission
of NGC~1569 is dominated by the high-column-den\-sity ridge and the western arm.  
NGC~1569-HI is $8'$ east of the centre of NGC~1569, 
and has a FWHM velocity width of $\rm 50\ km\ s^{-1}$. 
The spatially-integrated emission of NGC~1569-HI and of the bridge over 
the velocity range $\rm -190\ km\ s^{-1}$ to $\rm -60\ km\ s^{-1}$, is 
12.4 $\rm Jy\ km\ s^{-1}$, corrected for 16$\%$ primary-beam attenuation. 
The line flux of NGC 1569-HI itself is 6.3 $\rm Jy\ km\ s^{-1}$, half
of the total. If they are at a distance of 2.2 Mpc, the masses of NGC~1569-HI 
and of the bridge are $\rm 7 \cdot 10^6\ M_\odot$ each. 

\begin{figure}
\resizebox{\hsize}{!}{\includegraphics[angle=-90]{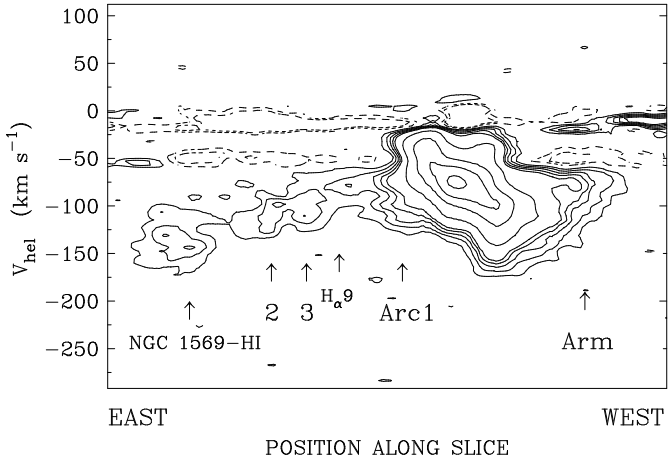}}
\caption{Position-velocity map along the HI bridge, the ridge, and the arm
extracted from the $60''$ resolution data cube. Specific positions mentioned in 
the text have been idicated. Contour levels : -10,-5,5,10,15,20,25,50,100,150,200,250 
mJy/beam per 4.12 $\rm km\ s^{-1}$ channel. Negative contours are drawn as dashed lines.}
\label{bridgeXV}
\end{figure}

On the side of NGC~1569, the apparent connection with the bridge is located
on the southern side of the galaxy. This region coincides with
the location of two concentric $H_\alpha$ shells described by Waller (1991).  

Figure~\ref{bridgeXV} shows a position velocity map along the curved dashed line
in Fig.~\ref{profig}. Going westwards from NGC 1569-HI, `bridge' emission occurs at 
progressively less negative velocities. 
The rotation of NGC 1569 occurs in the {\it opposite} sense. 

We have scanned the red and blue POSS II plates on the Leiden ASTROSCAN,
but failed to find an optical counterpart of NGC 1569-HI. 
The detection limit on the blue plates
for a smooth disk of $1'$ diameter should be about 26 mag arcsec$\rm ^{-2}$; 
foreground stellar confusion may raise this somewhat. 
With 2.3 magnitudes foreground extinction (Israel 1988), we  
arrive at a limit of about 23.5 mag arcsec$\rm ^{-2}$. As this is 
close to the median surface 
brightness of DDO dwarf galaxies (de Vaucouleurs et al. 1981), 
the optical non-detection of NGC 1569-HI therefore allows it to be
a low-surface-brightness dwarf galaxy with a foreground extinction
similar to that of NGC 1569, an intergalactic HI cloud without
any stars, or a Galactic high-velocity cloud.

\begin{figure}
\resizebox{8.0cm}{!}{\includegraphics{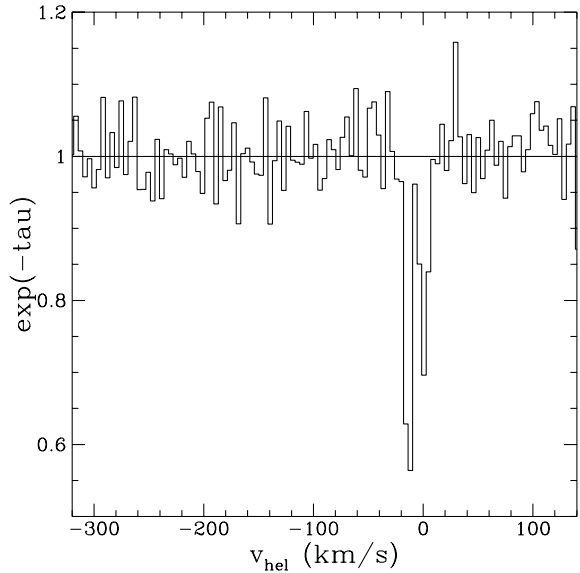}}
\caption{ Spectrum of background continuum sources 
normalized to the mean of 91 mJy between $\rm -300\ km s^{-1}$ and  $\rm -60\ km s^{-1}$;
rms fluctuations are 4$\%$. The brightest continuum source contributes 
approximately 50 $\%$ of the flux. Baselines shorter than 300 m have been 
excluded in order to suppress foreground emission. No velocity smoothing has 
been applied.}
\label{contabs}
\end{figure}

The central velocity of NGC~1569-HI is $\rm -130 \pm 5\ km\ s^{-1}$, which
is in the extreme blue wing of the line-of-sight Galactic HI line profile
(Hartmann \& Burton 1997). The Galactic foreground is 
clearly present at $v \rm > -70\ km\ s^{-1}$ throughout 
the field of view, but with a velocity width less than $\rm 10\ km\ s^{-1}$.
However, we find no emission other than that of NGC 1569 and NGC 1569-HI at 
any velocity $v_{hel} \rm  < -120\ km\ s^{-1}$, even at $60''$ resolution. 
We have summed the spectra of background point sources to look for absorption
(Fig.~\ref{contabs}). The $3\ \sigma$ upper limit to the optical depth per
channel is 0.12 in the velocity range $\rm -300\ km\ s^{-1}$ to $\rm -30\ km\ s^{-1}$. 
The corresponding upper limit to the HI column-density per channel of an
extended component with spin temperature $\rm T_s$, is $\rm N_{HI} < 9.6 \cdot 10^{19} (T_s/100 K)$. 

\section{Discussion.}

We first consider whether a Galactic origin is 
consistent with both the large velocity range of the eastern emission and its small 
angular size. The high-velocity-cloud complex closest to NGC 1569 is complex H, at a
distance of 5 degrees (Wakker 1991) and is in  
the relevant velocity range of $\rm -120\ km s^{-1}$ to $\rm -140\ km s^{-1}$. 
In Table~\ref{compare-tab} we compare the properties of NGC 1569-HI to those of 
HVC cores, the dwarf galaxy M81 dwA, and a sample of dwarf galaxies. 

No distinction can be made from the column density. 
However, the linewidth of NGC~1569-HI is much higher than expected for
small-scale structure in the Galactic foreground. The timescale for
free expansion to double its present size is short, $2 \cdot 10^4 \rm d_{kpc}$
year, where $\rm d_{kpc}$ is the distance in kpc.
Together with the lack of other isolated
HI clouds in the field of view, the apparent connection (bridge) with NGC 1569, and
the relatively high contrast with the surroundings, leads us to
conclude that NGC~1569-HI is not a condensation in the Galactic foreground.

Although NGC~1569 shows the signs of a global outflow of gas 
(see e.g. Heckman et al. 1995), the mass of 
NGC~1569-HI and the bridge ($\rm 7 \cdot 10^6\ M_\odot$ each) as well as their 
location projected along the major axis all rule out that they result from
this outflow. 

Extended, low-column-density HI haloes have been detec\-ted in a
number of late type galaxies (Huchtmeier \& Richter 1982). NGC~1569-HI
and the western arm are possibly high density regions in a similar
halo. The only evidence for extended emission with a column density
$\rm N_{HI} \sim 10^{20} cm^{-2}$ from the present data is the
emission in the north-west of NGC~1569. Its possible counterpart, symmetric in
position and in velocity with respect to the centre of NGC~1569 should
appear in the channel maps near $-60 \rm km\ s^{-1}$. Negative
sidelobes of Galactic foreground emission inhibit the detection of an
extended, low-sur\-face-brightness component (see e.g. Fig.~\ref{bridgeXV}).  
NGC~1569-HI cannot be
identified with the symmetric counterpart of the extended emission in
the northwest, because its velocity is $80$ $\rm km\ s^{-1}$ lower.  An
extended HI halo is sometimes inferred from the difference in flux
measured by an interferometer and a single dish radio
telescope. However, single dish observations are heavily contaminated
by the large column of Galactic HI. Furthermore, there is no evidence
for an HI halo in the maps presented by Reakes (1980).

The {\it apparent} retrograde rotation of NGC~1569-HI does not exclude
a prograde orbit, e.g. as a part of an extended HI disk.  
High resolution observations of the haloes of IC~10 (Shostak \&
Skillman 1989) and NGC~4449 (Bajaja et al. 1994) show a reversal of the
velocity field.  Such a reversal can be the result of a warped HI disk
bending through the plane of the sky. However, the warp angle must
then be larger than the inclination angle, which is $\rm 60^o$ for NGC~1569
(Israel 1988). There is no indication from the HI distribution that such an extreme
warp exists in NGC~1569.

As it is unlikely that NGC~1569-HI is an extension of the rotating disk of 
NGC~1569, it is either primordial gas which has not yet fallen into NGC~1569,
or a separate object in a close encounter with NGC~1569. The difference between 
these interpretations is mainly semantic.
No definite choice can be made since there is no optical 
identification of NGC~1569-HI. However, some general remarks can be made.

If NGC~1569-HI is a part of an HI halo which settles into the inner disk
of NGC~1569 on a cosmic timescale, its small distance to NGC~1569 suggests
that its is slowly spiraling inwards. 
This implies a relatively circular orbit, as opposed to a parabolic or
hyperbolic orbit expected for an encounter. Assuming a circular orbit,
a lower limit to the mass of NGC~1569 can be derived.  As the
projected distance, $d$, and radial velocity difference $\Delta v$ are
lower limits to the actual distance and relative velocity, the minimum
mass of the system if it is to be bound, must be 
$M_{min} \approx G^{-1} d \Delta v^2 = 2 \cdot 10^9 \rm M_\odot$.  
This lower limit, which is absolute for a circular orbit, is an order 
of magnitude higher than the estimated {\it dynamical} mass of NGC~1569 
within a radius of 1.3 kpc, where the rotation velocity is maximal (Reakes 1980).  
The lower limit to the total mass
results in a strong upper limit to the HI mass to total mass ratio
$M_{HI}/M_{total} < 0.04$ within 5 kpc, as opposed to the observed
$M_{HI}/M_{total} = 0.33$ within a radius of 1.3 kpc. 
The upper limit for the HI mass to total mass ratio is comparable to the ratio 
in dark matter dominated dwarf irregular galaxies such as DDO~154 
(Carignan \& Beaulieu 1989) at the last measured point
of the rotation curve. However, the HI mass to total mass ratio
is relatively constant with radius because the HI partial rotation
curve mimics the total rotation curve to a high accuracy
(Carignan \& Beaulieu 1989, Broeils 1992).
The combination $M_{HI}/M_{total} = 0.33$ within 1.3 kpc and $M_{HI}/M_{total} < 0.04$ 
within 5 kpc, is exceptional. This contradiction
is solved if the system is not gravitationally bound.

The velocity difference between NGC~1569 and NGC~1569-HI is 40 $\rm km\ s^{-1}$, 
their projected separation is 5 kpc, and their HI mass ratio
is about 0.06 (excluding the bridge). Corresponding values for other
pairs of dwarf galaxies studied by Taylor et al. (1995) are $\rm 50\ km\ s^{-1}$, 
30 kpc, and 0.2 respectively. Thus, NGC 1569 and NGC~1569-HI appear to be 
a close, but not an exceptional pair.

The virial mass of NGC~1569-HI is 
$\rm M_{vir} \sim R_{HI} \Delta \it v^2/G = \rm 6 \cdot 10^8\ M_\odot$.
If the HI mass ratio is representative for the total mass ratio, the ratio of
the tidal radius to their actual separation is 0.4. Judging from
their projected separation, NGC~1569-HI almost fills its tidal
radius. Tidal forces exerted by NGC~1569 may thus explain the
HI bridge. However, the western arm is unlikely to be the result of
tidal forces exerted by NGC~1569-HI given the small mass and the apparent 
retrograde orbit of NGC~1569-HI. This leaves the western arm unexplained.
If both are part of an outer halo which is not in equilibrium, as was 
suggested for IC~10, this problem does not exist.

It is evident from Fig.~\ref{bridgeXV} that the velocity structure
of NGC 1569-HI and the bridge is well ordered. It is therefore a
coherent structure and not a chance superposition of multiple clouds.
The timescale on which this structure evolves is in the order 
$d/\Delta v = 10^8 \rm years$. It is therefore a transient feature. 
Shostak \& Skillman (1989) found a structure similar to the HI bridge
and a disturbed location where this ``plume'' meets the disk of IC 10.
This was used as an argument in favour of interaction. A similar
argument can be made for NGC~1569.

The present data do not unequivocally show that the bridge is falling into
NGC 1569, but we may note some indications that this is so. First, the
presumed contact region exhibits chaotic kinematics (see Heckman et al. 1995 for
$\rm H_\alpha$ and our forthcoming paper for HI kinematics). Second, the
$\rm H_\alpha$ arcs and the extended X-ray feature described by 
Heckman et al. (1995) are found in the same direction. However, $\rm H_\alpha$ 
filaments are common in starburst galaxies, while a galactic wind is still 
needed to obtain the high Mach number required to heat the gas 
to X-ray temperatures. Finally, we note that the centre of the 
starburst is offset to the south from the high column density HI ridge,
i.e. towards the bridge. It is tempting to speculate
about a possible connection between recent bursts of star formation and
infalling material.

\begin{table}
\caption{Physical parameters of NGC 1569-HI compared with HVC cores
and low-mass dwarf galaxies.}

\tabcolsep=1.5mm
\begin{tabular}{|  l    l   l   l  l |} 
\hline 
                               &   N1569HI$^{\rm a}$ &  HVC core$^{\rm b}$  &  M81 dwA$^{\rm c}$ & LSY dw.$^{\rm d}$\\ 
\hline 
$\rm d_{Mpc}$   &   $\rm 2.2 $        & $\rm d_{Mpc}$      &  $\rm  3.25 $ & 0.76 - 5\\ 
$\rm D_{HI}$ (kpc)   &  $\rm 2.6 \times 1.3$     &  $\rm 0.6 \ d_{Mpc}$ &  $\rm 1.9 \times 1.3$ & 0.8 - 3.4\\ 
$\rm N_{HI,p}$  &$\rm 3.5 \cdot 10^{20}$&$\rm 3 \cdot 10^{20}$& $\rm 4 \cdot 10^{20}$ & 2.5 - 42 \\ 
$\rm \Delta v$ &  50              &  5               &  19  & 20 - 36 \\ 
$\rm \mu_B$   & $\rm > 23.5$       &  -            & 24.4 & 23.4 - 25.2 \\ 
\hline 
\end{tabular}
\begin{minipage}[b]{\columnwidth}
\footnotesize
Note : $\rm d_{Mpc}$ : distance in Mpc; $\rm D_{HI}$ : HI diameter (kpc); $\rm N_{HI,p}$ :
peak HI column density in $\rm 10^{20}\ HI\ cm^{-2}$; $\Delta v$ : HI FWHM line width 
($\rm km\ s^{-1}$); 
$\mu_B$ : mean surface brightness in B ($\rm mag\ arcsec^{-2}$);
References: {\it a.} This work; {\it b.} Wakker \& Schwarz (1991); {\it c.} Sargent et al. (1983);
{\it d.} range of the sample of 9 faint dwarfs in Lo et al. (1993). 
\end{minipage}
\normalsize 
\label{compare-tab}
\end{table}

\acknowledgements{We wish to thank R. Sancisi, W.B. Burton and P. Van der Werf 
for their useful comments about this paper and E. Deul for his help with the astroscan.}

\end{document}